\documentclass{bmcart}

\usepackage[utf8]{inputenc}
\usepackage{graphicx, url, csquotes}
\usepackage{xcolor,colortbl}
\usepackage{todonotes}

\startlocaldefs
\endlocaldefs

\begin{document}

\begin{frontmatter}

\begin{fmbox}
\dochead{Research}


\title{MP Twitter Abuse in the Age of COVID-19: White Paper}


\author[
   addressref={aff1},                   
   corref={aff1},                       
   email={g.gorrell@sheffield.ac.uk}   
]{\inits{GM}\fnm{Genevieve} \snm{Gorrell}}
\author[
   addressref={aff2},
   email={tracie.farrell@open.ac.uk}
]{\inits{T}\fnm{Tracie} \snm{Farrell}}
\author[
   addressref={aff1},
   email={k.bontcheva@sheffield.ac.uk}
]{\inits{KL}\fnm{Kalina} \snm{Bontcheva}}


\address[id=aff1]{
  \orgname{Department of Computer Science, Sheffield University}, 
  \street{Regent Court, 211 Portobello},                     %
  \city{Sheffield},                              
  \cny{UK}                                    
}
\address[id=aff2]{
  \orgname{Knowledge Media Institute, The Open University}, 
  \street{Walton Hall},                     %
  \city{Milton Keynes},                              
  \cny{UK}                                    
}


\begin{artnotes}
\end{artnotes}

\end{fmbox}


\begin{abstractbox}

\begin{abstract} 

As COVID-19 sweeps the globe, outcomes depend on effective relationships between the public and decision-makers. In the UK there were uncivil tweets to MPs about perceived UK tardiness to go into lockdown. The pandemic has led to increased attention on ministers with a role in the crisis. However, generally this surge has been civil. Prime minister Boris Johnson's severe illness with COVID-19 resulted in an unusual peak of supportive responses on Twitter. Those who receive more COVID-19 mentions in their replies tend to receive \textit{less} abuse (significant negative correlation). Following Mr Johnson's recovery, with rising economic concerns and anger about lockdown violations by influential figures, abuse levels began to rise in May. 1,902 replies to MPs within the study period were found containing hashtags or terms that refute the existence of the virus (e.g. \#coronahoax, \#coronabollocks, 0.04\% of a total 4.7 million replies, or 9\% of the number of mentions of ``stay home save lives'' and variants). These have tended to be more abusive. Evidence of some members of the public believing in COVID-19 conspiracy theories was also found. Higher abuse levels were associated with hashtags blaming China for the pandemic.

\end{abstract}


\begin{keyword}
\kwd{COVID-19}
\kwd{Twitter}
\kwd{politics}
\kwd{incivility}
\kwd{abuse}
\end{keyword}


\end{abstractbox}
%

\end{frontmatter}



\section*{Introduction}

A successful response to the COVID-19 pandemic depends on effective relationships between the public and decision-makers. Yet the pandemic arises in the midst of the age of misinformation~\cite{bessi2015science, vaezi2020infodemic},\footnote{\small{Page 2 ``infodemic'': \url{https://www.who.int/docs/default-source/coronaviruse/situation-reports/20200202-sitrep-13-ncov-v3.pdf}, \url{https://www.bbc.co.uk/news/technology-51497800}}}, creating a perfect storm. Polarisation and echo chambers make it harder for the right information to reach people, and for them to trust it when it does~\cite{zollo2015emotional}, and the damage can be counted in lives.\footnote{\small{\url{https://www.bbc.co.uk/news/stories-52731624}, \url{https://www.independent.co.uk/news/world/middle-east/iran-coronavirus-methanol-drink-cure-deaths-fake-a9429956.html}, \url{https://thehill.com/changing-america/well-being/prevention-cures/494548-maryland-emergency-hotline-receives-more-than}, \url{https://www.mirror.co.uk/news/us-news/coronavirus-man-dies-after-drinking-21743881}}} Online verbal abuse is an intrinsic aspect of the misinformation picture, being both cause and consequence: the quality of information and debate is damaged as certain voices are silenced/driven out of the space,\footnote{\small{\url{https://www.bbc.co.uk/news/election-2019-50246969}}} and the escalation of divisive ``outrage'' culture leads to angry and aggressive expressions~\cite{sobieraj2011incivility}.

This white paper charts Twitter abuse in replies to UK MPs, and a number of other prominent/relevant accounts, from before the start of the pandemic in the UK until late May, in order to plot the health of relationships of UK citizens with their elected representatives through the unprecedented challenges of the COVID-19 epidemic. We consider reactions to different individuals and members of different political parties, and how they interact with events relating to the virus. We review the dominant hashtags on Twitter as the country moves through different phases. The six periods considered are as follows:

\begin{itemize}
\item February 7th to 29th inclusive: Control: little attention to COVID-19
\item March 1st to 22nd inclusive: Growing awareness of COVID-19, culminating in the week in which we were advised but not yet obliged to begin social distancing
\item March 23rd to March 31st: Beginning of lockdown
\item April 1st to April 16th: Middle-lockdown
\item April 17th to May 9th: Emergence of global backlash against lockdown
\item May 10th to May 25th: Easing of lockdown
\end{itemize}

We show trends in abuse levels, for MPs overall as well as for particular individuals and for parties. We conclude with a section that compares prevalence of conspiracy theories, and contextualises them against other popular topics/concerns on Twitter. We begin with a brief summary of related work, before outlining the methodology used and then progressing onto findings.

\section*{Related work}

A raft of work has rallied to focus attention on COVID-19, as the scientific community recognises the extent to which outcomes depend on information and strategy. Research has begun to address the role of internet and social media in development of attitudes, compliance, and adoption of effective responses, and the way misinformation can derail these~\cite{cuan2020misinformation, kouzy2020coronavirus, freeman2020coronavirus}. Furthermore, as the pandemic increasingly puts pressure on the divisions in society, with mortality risk much greater for some communities than others,\footnote{\small{\url{https://www.theguardian.com/world/2020/may/07/black-people-four-times-more-} \url{likely-to-die-from-covid-19-ons-finds}}} we are forced to recognise the pandemic as a highly political issue (e.g. Motta et al~\cite{motta2020right}). Pew\footnote{\small{\url{https://www.pewresearch.org/pathways-2020/COVIDCREATE/main_source_of_election_news/us_adults}}} repeatedly find attitudes toward the disease split along partisan lines. In the age of COVID-19, polarisation can be deadly.

Twitter has become a favoured platform for politicians across the globe, providing a means by which the public can communicate directly with them. Previous work~\cite{gorrell2018twits, gorrell2019race, binns2018and, ward2017turds} has shown rising levels of hostility towards UK politicians on Twitter in the context of divisive issues, and we also see partisan operators seeking to gain influence in the space~\cite{gorrell2019partisanship}. Women and minorities have been shown to have different and potentially more threatening online experiences, raising concerns about representation~\cite{gorrell2019race, delisle2019large, pew2017}. Moving into the COVID-19 era, key questions raised therefore are about the impact of the pandemic on hostility levels towards decision-makers, the opportunities it presents for partisan operators to further damage social cohesion, and the impact on effectiveness and experience of women and minority politicians. This work focuses on the first two of these subjects.

\section*{Methodology}

In this work we utilize a large tweet collection on which a natural language processing has been performed in order to identify abusive language. This methodology is presented in detail by Gorrell \textit{et al}~\cite{gorrell2019race} and summarised here.

A rule-based approach was used to detect abusive language. An extensive vocabulary list of slurs (e.g. ``idiot''), offensive words such as the ``f'' word and potentially sensitive identity markers, such as ``lesbian'' or ``Muslim'', forms the basis of the approach. The slur list contained 1081 abusive terms or short phrases in British and American English, comprising mostly an extensive collection of insults, racist and homophobic slurs, as well as terms that denigrate a person's appearance or intelligence, gathered from sources that include \url{http://hatebase.org} and Farrell \textit{et al} \cite{farrell2019exploring}. 131 offensive worrds were used, along with 451 sensitive words. ``Bleeped'' versions such as ``f**k'' are also included.

On top of these word lists, 53 rules are layered, specifying how they may be combined to form an abusive utterance as described above, and including further specifications such as how to mark quoted abuse, how to type abuse as sexist or racist, including more complex cases such as ``stupid Jew hater'' and what phrases to veto, for example ``polish a turd'' and ``witch hunt''. Making the approach more precise as to target (whether the abuse is aimed at the politician being replied to or some third party) was achieved by rules based on pronoun co-occurrence. The approach is generally successful, but where people make a lot of derogatory comments about a third party in their replies to a politician, for example racist remarks about others, there may be a substantial number of false positives.

The abuse detection method underestimates by possibly as much as a factor of two, finding more obvious verbal abuse, but missing linguistically subtler examples. This is useful for comparative findings, tracking abuse trends, and for approximation of actual abuse levels.

The method for detecting COVID-19-related tweets is based on a list of related terms. This means that tweets that are implicitly about the epidemic but use no explicit Covid terms, for example, ``@BorisJohnson you need to act now,'' are not flagged. The methodology is useful for comparative findings such as who is receiving the most COVID-19-related tweets, but not for drawing conclusions about absolute quantities of tweets on that subject.

\section*{Corpus}

The corpus was created by collecting tweets in real-time using Twitter's streaming API. We used the API to follow the accounts of MPs - this means we collected all the tweets sent by each candidate, any replies to those tweets, and any retweets either made by the candidate or of the candidate's own tweets. Note that this approach does not collect all tweets which an individual would see in their timeline, as it does not include those in which they are just mentioned. However, ``direct replies'' are included. We took this approach as the analysis results are more reliable  due to the fact that replies are directed at the politician who authored the tweet, and thus, any abusive language is more likely to be directed at them. Data were of a low enough volume not to be constrained by Twitter rate limits.

The study spans February 7th until May 25th 2020 inclusive, and discusses Twitter replies to currently serving MPs that have active Twitter accounts (568 MPs in total). Table 1 gives the overall statistics for the corpus. Dates in the table indicate a period from midnight to midnight at the start of the given date.

Tweets from earlier in the study have had more time to gather replies. Most replies occur in the day or two following the tweet being made, but some tweets continue to receive attention over time, and events may lead to a resurgence of focus on an earlier tweet. Reply numbers are a snapshot at the time of the study.

\begin{table}
\begin{tabular}{lrrrrrr}
Time period & Original & Retweet & Reply & ReplyTo & Abusive & \% Abuse\\
\hline
02/07-03/01 & 16,482 & 26,632 & 6,952 & 562,322 & 19,301 & 3.432\\
03/01-03/23 & 22,419 & 39,781 & 11,482 & 777,396 & 33,069 & 4.254\\
03/23-04/01 & 11,571 & 21,821 & 7,137 & 441,983 & 13,919 & 3.149\\
04/01-04/17 & 17,007 & 30,124 & 10,407 & 782,774 & 24,327 & 3.108\\
04/17-05/10 & 22,906 & 38,949 & 11,906 & 890,926 & 32,050 & 3.597\\
05/10-05/26 & 16,824 & 30,279 & 8,822 & 1,270,669 & 56,827 & 4.472\\
\hline
Total & 107,209 & 187,586 & 56,706 & 4,726,070 & 179,493 & 3.798\\
\end{tabular}
\caption{Corpus statistics. Columns give, for each time period, number of original tweets authored by MPs, number of retweets authored by them, number of replies written by them, number of replies received by them, number of abusive replies received by them, and abusive replies received as a percentage of all replies received.}
\label{tab:corpus}
\end{table}

\section*{Findings}

We begin with a review of the time period studied, namely February 7th until May 25th inclusive. After that, findings are organised into time periods with distinct characteristics with regards to the course of the pandemic in the UK, as listed in the introduction. We then include a section on conspiracy theories.

\subsection*{Overview}

\begin{figure}
  \includegraphics[width=.85\textwidth]{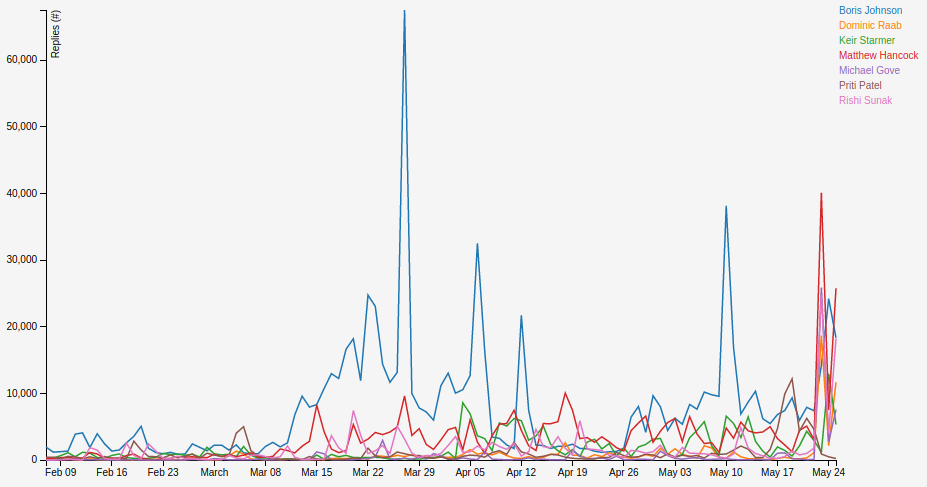}
  \caption{Number of replies received by relevant ministers and opposition leader Keir Starmer per day from February 7th to May 25th inclusive.}
  \label{fig:timeline}
\end{figure}

\begin{figure}
  \includegraphics[width=.85\textwidth]{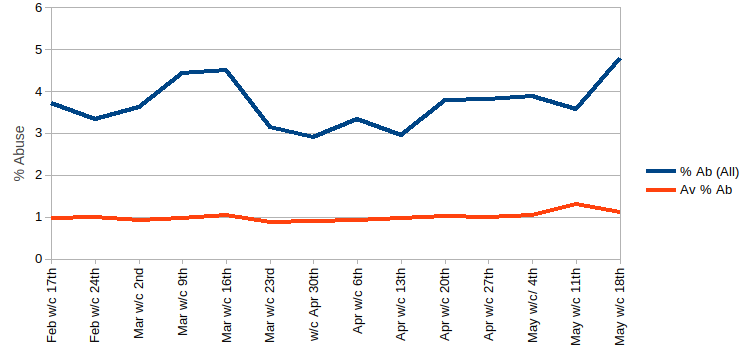}
  \caption{Abuse percentage received by all MPs, macro- (red) and micro- (blue) average, per week}
  \label{fig:timeline-per-week}
\end{figure}

Fig~\ref{fig:timeline} shows number of replies to prominent politicians since early February, and shows that for the most part, attention has focused on Boris Johnson. He received a large peak in Twitter attention on March 27th. 58,286 replies were received in response to his tweet announcing that he had COVID-19. Abuse was found in 2.3\% of these replies, which is low for a prominent minister as we may discern from Fig~\ref{fig:timeline-per-week}, suggesting a generally supportive response to the prime minister's illness. Further peaks on Mr Johnson's timeline correspond to the dates on which he was admitted to intensive care (April 6th), left hospital to recuperate at Chequers (April 12th), and more recently, began to ease the lockdown (May 10th). The late burst of attention on other politicians arises from several tweets by ministers in support of Dominic Cummings, the senior government advisor who chose to travel north to his parents' home in the early stages of his illness with COVID-19.

The timeline in Fig~\ref{fig:timeline-per-week} makes it easier to see abuse levels overall, toward all MPs. It is on a per-week basis since mid-February, and shows a rise in abuse, back up to over 4\% around the time of the introduction of social distancing, before dipping, and then gradually beginning to rise again in recent times. The dip may be explained, in part at least, by an unusual degree of positive attention focused on the prime minister as he faced personal adversity, depressing the abuse level as a percentage of all replies. We see that the macro-average abuse level (red line) remains relatively steady.

\subsection*{Difference in responses to different parties}

\begin{figure}
  \includegraphics[width=.85\textwidth]{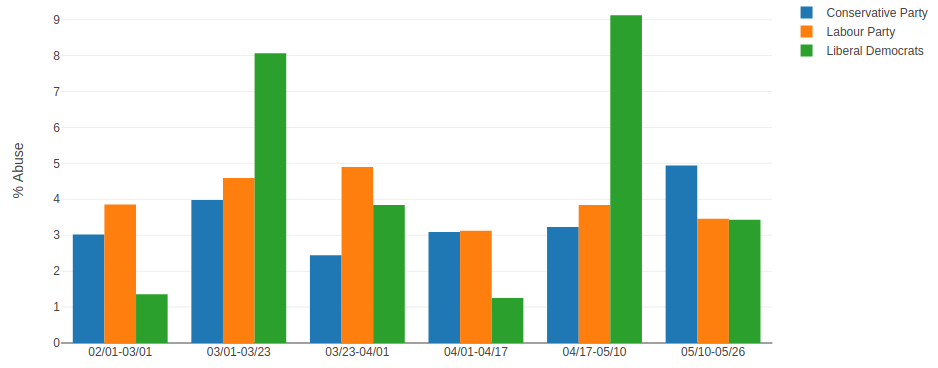}
  \caption{Abusive replies as a percentage of all replies received, micro-average, split by party and time period.}
  \label{fig:parties}
\end{figure}

Fig~\ref{fig:parties} shows abuse received as a percentage of all replies received by MPs, for each of the time periods studied in more detail below. We see that on the whole, response to the Conservative party has been favourable, as indicated above. The exception is after May 10th, when the negative response to Dominic Cummings' decision to travel north with COVID-19 symptoms came to the fore. Responses to Liberal Democrat MPs are more erratic due to their lower number.\footnote{\small{The peak in early March arises with a now-deleted tweet from Layla Moran. The peak in late April/early May arises with the start of Ramadan and a supportive tweet from Ed Davey.}} In previous studies, we have found Conservatives receiving higher abuse levels, yet here we see Labour politicians receiving more abuse in most periods. This was in evidence even in February, so precedes the pandemic. Twitter has tended to be left-leaning in the UK~\cite{gorrell2019local} - it remains to be seen if this is the beginning of a swing to the right or if it is specific to the times, e.g. arising from a desire to trust authority during times of crisis \cite{volkan2014blind}.

There is a significant negative correlation between receiving a high level of Covid-related attention and receiving abuse (-0.52, p<0.001, Feb 7th to May 25th, Spearman's PMCC). We see this clearly in prominent government figures below, who are receiving the lion's share of the COVID-19 attention and lower levels of abuse than we usually see for them. However the correlation is significant across the sample of all MPs, suggesting perhaps that an association with the crisis is generally good for the image.

\subsection*{Long-term context}

\begin{figure}
  \includegraphics[width=.85\textwidth]{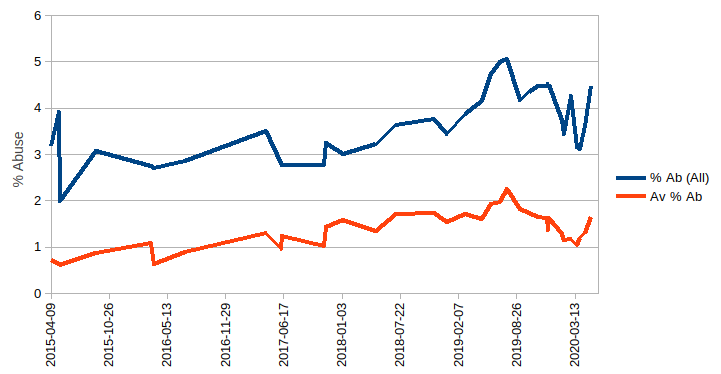}
  \caption{Timeline of abuse received by MPs from 2015 until mid-February 2020. Blue line is micro-average and red line is macro-average.}
  \label{fig:timeline-long}
\end{figure}

From the timeline in Fig~\ref{fig:timeline-long}, we see that aside from a blip around the 2015 general election, abuse toward MPs on Twitter has been tending to rise from a minimum of 2\% of replies in 2015, peaking mid-2019 at over 5\% with a smaller peak of around 4.5\% around the 2019 general election. After the election, however, abuse toward MPs fell to around 3.5\%.\footnote{\small{Due to the variable quality nature of the historical data, the time periods across which these data points are averaged varies, which may lead to shorter-term peaks being lost. We hope to improve on this in future work. Where more datapoints are available, the graph appears more spiky, as we see with the richer recent data.}} A spike followed before the beginning of lockdown, before a low, and more recently, a rise, as discussed above. However the low was not as low as in 2015, and the high is not as high as the Brexit acrimony of 2019. The blue line shows micro-averaged abuse as a percentage of replies. The red line shows macro-averaged abuse percentage (percentage is calculated per person, then averaged). Where the two lines differ, we can infer an unusual response at that time particularly to high-profile politicians.

\subsection*{February 7th to 29th}

Table~\ref{tab:feb-7} provides some quick reference information for the top 20 MPs receiving the largest number of replies to their Twitter account during the February period. The column “Authored” refers to the number of tweets originally posted from that account that were not retweets or replies. ``replyTo'', refers to all of the replies received to the individual’s Twitter account in that period. The next column, ``Covid'', is the number of replies received to that account containing an explicit mention of COVID-19, with the following column representing the number of replies that verbal abuse was found in (``Abusive''). The last three columns present the data in a comparative fashion. Firstly, we have the percentage of replies that the individual received that were abusive. Next, we have the percentage of replies that were Covid related. The last column is the percentage of Covid-related replies to that individual, in comparison with all Covid-related replies received by all MPs. We have created a table and histogram for each period.

\begin{table}
\begin{tabular}{lrrrrrrr}
Name & Authored & replyTo & Covid & Abusive & \% Ab & \% Covid & Total Covid\\
\hline
\cellcolor{blue!15}Boris Johnson & 14 & 48,379 & 1,072 & 1,695 & 3.504 & 2.216 & 37.773\\
\cellcolor{red!15}David Lammy & 89 & 47,368 & 73 & 2,308 & 4.872 & 0.154 & 2.572\\
\cellcolor{red!15}Richard Burgon & 184 & 30,789 & 18 & 1,556 & 5.054 & 0.058 & 0.634\\
\cellcolor{red!15}Jeremy Corbyn & 25 & 29,550 & 215 & 2,400 & 8.122 & 0.728 & 7.576\\
\cellcolor{red!15}Rebecca Long-Bailey & 121 & 27,113 & 17 & 823 & 3.035 & 0.063 & 0.599\\
\cellcolor{red!15}Zarah Sultana & 96 & 18,630 & 12 & 677 & 3.634 & 0.064 & 0.423\\
\cellcolor{red!15}Debbie Abrahams & 96 & 18,186 & 45 & 822 & 4.520 & 0.247 & 1.586\\
\cellcolor{red!15}Keir Starmer & 97 & 15,455 & 26 & 360 & 2.329 & 0.168 & 0.916\\
\cellcolor{red!15}Lisa Nandy & 110 & 15,271 & 9 & 413 & 2.704 & 0.059 & 0.317\\
\cellcolor{blue!15}Priti Patel & 9 & 13,664 & 45 & 345 & 2.525 & 0.329 & 1.586\\
\cellcolor{blue!15}Rishi Sunak & 10 & 12,674 & 19 & 399 & 3.148 & 0.150 & 0.669\\
\cellcolor{blue!15}Sajid Javid & 26 & 10,945 & 14 & 362 & 3.307 & 0.128 & 0.493\\
\cellcolor{blue!15}Jacob Rees-Mogg & 13 & 10,352 & 22 & 584 & 5.641 & 0.213 & 0.775\\
\cellcolor{red!15}Jess Phillips & 106 & 9,648 & 0 & 140 & 1.451 & 0.000 & 0.000\\
\cellcolor{red!15}Dawn Butler & 152 & 9,457 & 3 & 231 & 2.443 & 0.032 & 0.106\\
\end{tabular}
\caption{MPs with greatest number of replies from February 7 - 29 2020 inclusive. Cell colours indicate party membership; blue for Conservative, red for Labour.}
\label{tab:feb-7}
\end{table}

\begin{figure}
  \includegraphics[width=.85\textwidth]{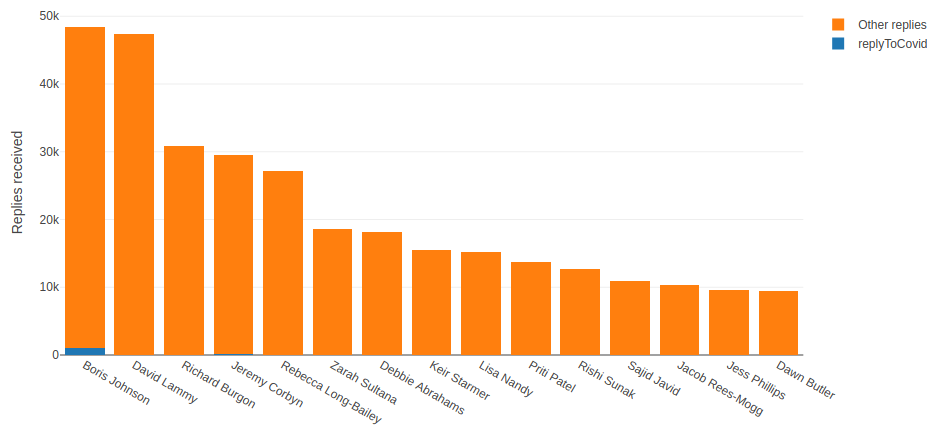}
  \caption{Number of replies with an explicit COVID-19 mention as a portion of all replies from February 7 - 29 2020 inclusive.}
  \label{fig:feb-7-histo}
\end{figure}

We see from Table~\ref{tab:feb-7} that Boris Johnson receives the most tweets, with controversial Labour politicians not far behind. Whilst he also receives the most abuse by volume, as a percentage of replies received, the 3.5\% shown here is unusually low for Mr Johnson compared with our findings in earlier work (e.g. 8.39\%\footnote{\small{https://arxiv.org/pdf/1910.00920.pdf}} in the first half of 2019). The histogram in~\ref{fig:feb-7-histo} shows the number of replies received related to COVID-19, in comparison with the number of replies received in general for that period. Once again, this chart indicates that attention to COVID-19 was limited at the start of the pandemic, but what attention there was to the subject was largely aimed at Mr Johnson.

\subsubsection*{Hashtags}

\begin{figure}
  \includegraphics[width=.85\textwidth]{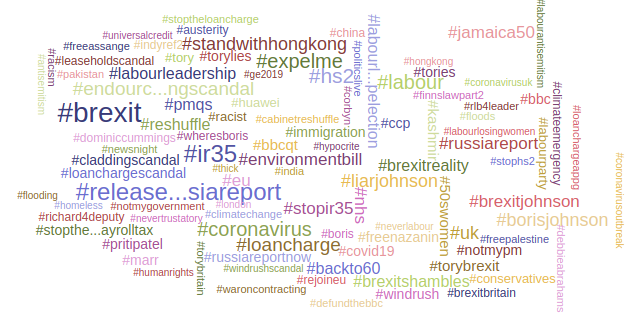}
  \caption{Top 100 hashtags in all replies sent to MPs - February 7-29 2020 inclusive.}
  \label{fig:feb-7-cloud}
\end{figure}

The hashtag cloud in~\ref{fig:feb-7-cloud} above shows that in February, little attention was focused on COVID-19, and Brexit remained the dominant topic in Twitter political discourse. Table~\ref{tab:feb-7-hash} gives the top ten hashtags in numeric terms.

\begin{table}
\centering
\begin{tabular}{lr}
Hashtag & Count\\
\hline
\#brexit & 3,955\\
\#ir35 & 1,964\\
\#releasetherussiareport & 1,688\\
\#hs2 & 1,099\\
\#expelme & 877\\
\#coronavirus & 831\\
\#labour & 810\\
\#endourcladdingscandal & 717\\
\#loancharge & 691\\
\#uk & 613\\
\end{tabular}
\caption{Hashtag counts in replies to MPs, February 7-29 inclusive.}
\label{tab:feb-7-hash}
\end{table}

\subsubsection*{Examples of tweets that attracted particularly abusive responses}

In February, COVID-19 was not a major focus. The most abused tweet of the month was this one, by then-leader of the opposition Jeremy Corbyn, attracting 4\% of all abusive replies to MPs.

\url{https://twitter.com/jeremycorbyn/status/1227589027790016514} (11\% abuse, 4\% of all abusive replies to MPs in February):

\begin{displayquote}
If there was a case of a young white boy with blonde hair, who later dabbled in class A drugs and conspired with a friend to beat up a journalist, would they deport that boy?

Or is it one rule for young black boys born in the Caribbean, and another for white boys born in the US?
\end{displayquote}

\subsection*{March 1st to 22nd}

In this section we review data from March 1st to March 22nd inclusive, when Boris Johnson made the first announcement that citizens were advised to avoid non-essential contact and journeys.

\begin{table}
\begin{tabular}{lrrrrrrr}
Name & Authored & replyTo & Covid & Abusive & \% Ab & \% Covid & Total Covid\\
\hline
\cellcolor{blue!15}Boris Johnson & 45 & 160,356 & 28,818 & 7,684 & 4.792 & 17.971 & 49.380\\
\cellcolor{red!15}David Lammy & 120 & 43,386 & 1,602 & 2,952 & 6.804 & 3.692 & 2.745\\
\cellcolor{blue!15}Matthew Hancock & 100 & 42,520 & 7,167 & 1,800 & 4.233 & 16.856 & 12.281\\
\cellcolor{red!15}Jeremy Corbyn & 44 & 42,435 & 1,824 & 3,244 & 7.645 & 4.298 & 3.125\\
\cellcolor{blue!15}Rishi Sunak & 47 & 25,534 & 2,225 & 284 & 1.112 & 8.714 & 3.813\\
\cellcolor{blue!15}Nadine Dorries & 52 & 24,731 & 1,275 & 466 & 1.884 & 5.155 & 2.185\\
\cellcolor{red!15}Jess Phillips & 154 & 20,931 & 394 & 541 & 2.585 & 1.882 & 0.675\\
\cellcolor{red!15}Lisa Nandy & 74 & 20,355 & 234 & 925 & 4.544 & 1.150 & 0.401\\
\cellcolor{red!15}Richard Burgon & 145 & 20,281 & 446 & 1,222 & 6.025 & 2.199 & 0.764\\
\cellcolor{red!15}Zarah Sultana & 107 & 18,796 & 204 & 815 & 4.336 & 1.085 & 0.350\\
\cellcolor{blue!15}Priti Patel & 12 & 17,233 & 451 & 378 & 2.193 & 2.617 & 0.773\\
\cellcolor{blue!15}Pauline Latham & 6 & 12,540 & 204 & 1,049 & 8.365 & 1.627 & 0.350\\
\cellcolor{red!15}James Cleverly & 49 & 11,544 & 606 & 650 & 5.631 & 5.249 & 1.038\\
\cellcolor{yellow!40}Layla Moran & 130 & 11,456 & 479 & 976 & 8.520 & 4.181 & 0.821\\
\cellcolor{red!15}Keir Starmer & 39 & 11,217 & 385 & 299 & 2.666 & 3.432 & 0.660\\
\end{tabular}
\caption{MPs with greatest number of replies from from March 1st - 22nd 2020 inclusive. Cell colours indicate party membership; blue for Conservative, red for Labour, yellow for Liberal Democrat.}
\label{tab:mar-1}
\end{table}

\begin{figure}
  \includegraphics[width=.85\textwidth]{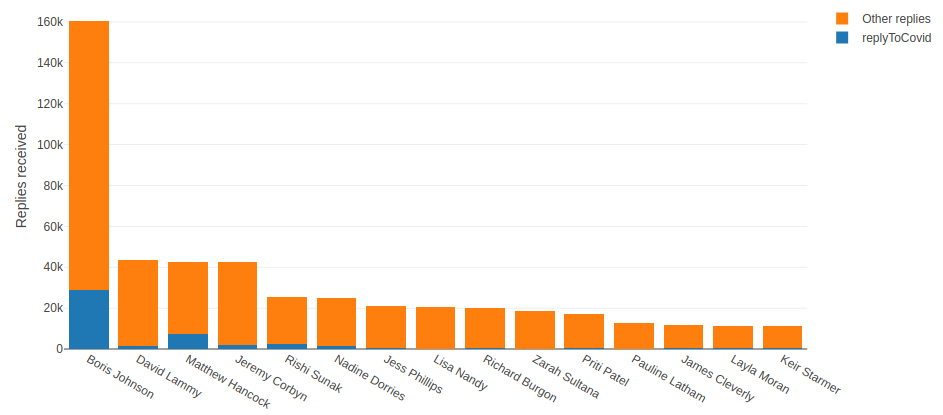}
  \caption{Number of replies with an explicit COVID-19 mention as a portion of all replies from March 1st - 22nd 2020 inclusive.}
  \label{fig:mar-1-histo}
\end{figure}

In Table~\ref{tab:mar-1} and Fig~\ref{fig:mar-1-histo}, one can see how attention on particular individuals has changed in the first period in March. In the table you can see the number of replies they receive, the percentage of those replies that are related to COVID-19, and how this compares with other MP colleagues. Health secretary Matt Hancock became more prominent on Twitter at this time, though attention was not more abusive. Attention on chancellor Rishi Sunak also increased and was not abusive. We see a high level of attention on Boris Johnson, but the abuse level is lower than was seen for him in previous years (we found 8.39\% in the first half of 2019 as mentioned above; in 2017 as foreign secretary Mr Johnson received similarly high abuse levels in high volumes). Negative attention on Labour politicians is high, but note that this was also the case before the start of the epidemic in the UK. A focus on COVID-19 is now in evidence (recall that counts for COVID-19 tweets only include explicit mentions; it is likely that many more replies are about COVID-19).

\subsubsection*{Hashtags}

\begin{figure}
  \includegraphics[width=.85\textwidth]{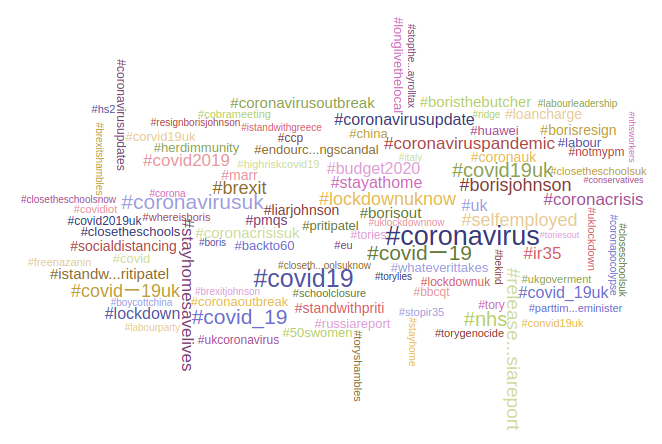}
  \caption{Top 100 hashtags in all replies sent to MPs - March 1st - 22nd 2020 inclusive.}
  \label{fig:mar-1-cloud}
\end{figure}

The word cloud in Fig~\ref{fig:mar-1-cloud} shows all hashtags in tweets to MPs in earlier part of March, and unsurprisingly shows a complete topic shift, to the subject of the epidemic, to the virtual exclusion of all else. The February word cloud shows a variety of non-Covid subjects, such as Brexit, IR35 (tax loophole legislation), the Russia report, climate change, pension age for women and the accusations against Priti Patel. Now, almost all hashtags are COVID-19-related. Table~\ref{tab:mar-1-hash} gives the counts for the top ten.\footnote{\small{A non-standard em dash was used in hashtags referring to COVID-19 for a time on Twitter - in the tables we show a standard em dash.}}

\begin{table}
\begin{tabular}{lr}
Hashtag & Count\\
\hline
\#coronavirus & 10,594\\
\#covid19 & 7,424\\
\#covid\_19 & 2,739\\
\#coronavirusuk & 2,613\\
\#covid19uk & 2,098\\
\#nhs & 1,925\\
\#releasetherussiareport & 1,717\\
\#covid---19 & 1,671\\
\#brexit & 1,519\\
\#covid---19uk & 1,494\\
\end{tabular}
\caption{Hashtag counts in replies to MPs, March 1st - 22nd inclusive}
\label{tab:mar-1-hash}
\end{table}

\subsubsection*{Examples of tweets that attracted particularly abusive responses}

As discussed above, feelings ran high around the beginning of lockdown, and examples of high volume tweets with elevated abuse levels were given. Here we give tweets receiving high abuse levels.

\url{https://twitter.com/BorisJohnson/status/1238365263764041728} (9\% of replies were abusive, tweet received 3\% of all abuse to MPs in the period). It also includes a video.

\begin{displayquote}
This country will get through this epidemic, just as it has got through many tougher experiences before.
\end{displayquote}

Another tweet that gradually rose to become the second most abused tweet by volume of the period was a now-deleted tweet by Layla Moran defending China, apropos COVID-19. 14\% abuse, 3\% of abuse for the period.

\url{https://twitter.com/Pauline_Latham/status/1238903676934160384} (12\% abuse, 2\% of all abuse to MPs in the March pre-lockdown period; told constituent to ``get a life'' in response to tweet about statutory sick pay).

\url{https://twitter.com/MattHancock/status/1238960146342084609} (11\% abuse, 3\% of all abuse to MPs in the March pre-lockdown period, inspiring critical responses):

\begin{displayquote}
NEWS: My Telegraph article on the next stage of our \#coronavirus plan:

We must all do everything in our power to protect lives
\end{displayquote}

\url{https://twitter.com/DavidLammy/status/1239835712444391424} (16\% abuse, 1\% of all abuse sent to MPs in the March pre-lockdown period):

\begin{displayquote}
No more government time, energy or resources should be wasted on Brexit this year. Boris Johnson must ask for an extension to the transition period immediately. \#COVID19 is a global emergency.
\end{displayquote}

\subsection*{March 23rd to 31st}

\begin{figure}
  \includegraphics[width=.85\textwidth]{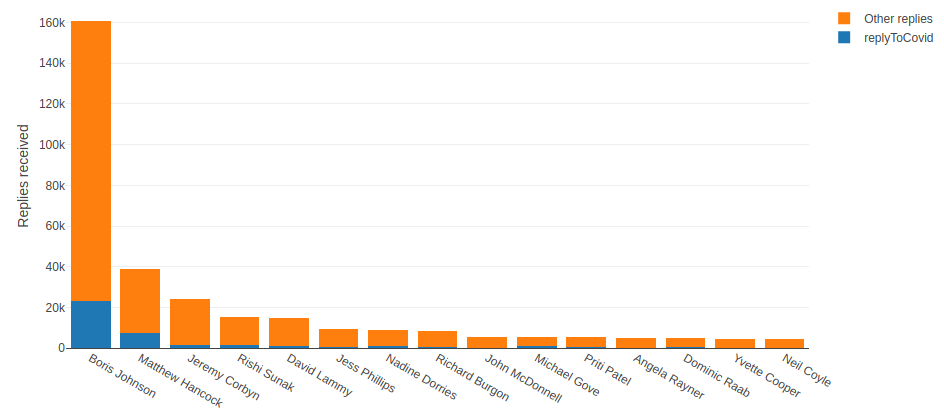}
  \caption{Number of replies with an explicit COVID-19 mention as a portion of all replies from March 22nd - 31st 2020 inclusive.}
  \label{fig:mar-23-histo}
\end{figure}

For the second period in March from the 23rd - 31st, attention on individual MPs was reshuffled relative to the number of replies they received. From Table~\ref{tab:mar-23} and Fig~\ref{fig:mar-23-histo} we can see that attention continues to focus on Boris Johnson, and is even less abusive than previously, largely due to a surge in non-abusive attention in conjunction with his being diagnosed with COVID-19. Matt Hancock becomes more prominent, though attracting less abuse than previously.

\begin{table}
\begin{tabular}{lrrrrrrr}
Name & Authored & replyTo & Covid & Abusive & \% Ab & \% Covid & Total Covid\\
\hline
\cellcolor{blue!15}Boris Johnson & 22 & 160,750 & 23,091 & 4,267 & 2.654 & 14.365 & 47.657\\
\cellcolor{blue!15}Matthew Hancock & 58 & 39,029 & 7,376 & 979 & 2.508 & 18.899 & 15.223\\
\cellcolor{red!15}Jeremy Corbyn & 24 & 23,952 & 1,378 & 1,867 & 7.795 & 5.753 & 2.844\\
\cellcolor{blue!15}Rishi Sunak & 13 & 15,451 & 1,557 & 139 & 0.900 & 10.077 & 3.213\\
\cellcolor{red!15}David Lammy & 69 & 14,608 & 853 & 656 & 4.491 & 5.839 & 1.761\\
\cellcolor{red!15}Jess Phillips & 100 & 9,648 & 345 & 162 & 1.679 & 3.576 & 0.712\\
\cellcolor{blue!15}Nadine Dorries & 40 & 9,041 & 920 & 315 & 3.484 & 10.176 & 1.899\\
\cellcolor{red!15}Richard Burgon & 47 & 8,253 & 569 & 941 & 11.402 & 6.894 & 1.174\\
\cellcolor{red!15}John McDonnell & 41 & 5,584 & 317 & 326 & 5.838 & 5.677 & 0.654\\
\cellcolor{blue!15}Priti Patel & 9 & 5,566 & 630 & 90 & 1.617 & 11.319 & 1.300\\
\cellcolor{blue!15}Michael Gove & 8 & 5,566 & 850 & 220 & 3.953 & 15.271 & 1.754\\
\cellcolor{red!15}Angela Rayner & 35 & 5,171 & 137 & 41 & 0.793 & 2.649 & 0.283\\
\cellcolor{blue!15}Dominic Raab & 26 & 4,984 & 410 & 47 & 0.943 & 8.226 & 0.846\\
\cellcolor{red!15}Yvette Cooper & 12 & 4,407 & 224 & 311 & 7.057 & 5.083 & 0.462\\
\cellcolor{red!15}Neil Coyle & 56 & 4,347 & 82 & 458 & 10.536 & 1.886 & 0.169\\
\end{tabular}
\caption{MPs with greatest number of replies from from March 23 -  31 2020 inclusive. Cell colours indicate party membership; blue for Conservative, red for Labour.}
\label{tab:mar-23}
\end{table}

\subsubsection*{Hashtags}

\begin{figure}
  \includegraphics[width=.85\textwidth]{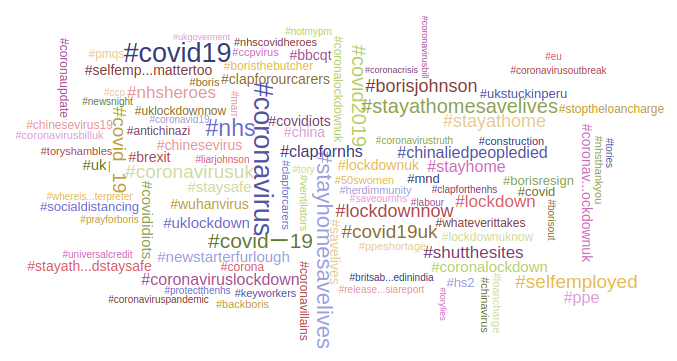}
  \caption{Top 100 hashtags in all replies sent to MPs - March 23rd - 31st 2020 inclusive.}
  \label{fig:mar-23-cloud}
\end{figure}

The rise of the hashtag ``\#stayhomesavelives'' shows a shift toward comment on the practical details (see Fig~\ref{fig:mar-23-cloud}). Support for the lockdown appears to be high at this stage. Table~\ref{tab:mar-23-hash} gives counts for the top ten.

\begin{table}
\begin{tabular}{lr}
Hashtag & Count\\
\hline
\#coronavirus & 7,593\\
\#covid19 & 7,527\\
\#stayhomesavelives & 3,324\\
\#nhs & 3,006\\
\#covid---19 & 2,917\\
\#lockdownuknow & 1,676\\
\#stayathomesavelives & 1,663\\
\#stayathome & 1,645\\
\#coronavirusuk & 1,473\\
\#covid\_19 & 1,463\\
\end{tabular}
\caption{Hashtag counts in replies to MPs, March 23rd - 31st.}
\label{tab:mar-23-hash}
\end{table}

\subsubsection*{Examples of tweets that attracted particularly abusive responses}

By volume, the most abused tweet was Boris Johnson's illness announcement, but as a percentage this was remarkably un-abusive, as discussed above, with 2.3\% abuse, and the abuse count follows only from the very high level of attention this tweet drew. The most striking tweet in terms of receiving a high percentage of abuse as well as a notable degree of attention was the one below from Richard Burgon.

\url{https://twitter.com/RichardBurgon/status/1244043022297370626} (17\% abuse, or 6\% of all abuse sent to MPs in March post-lockdown):

\begin{displayquote}
This is a Trump-style attempt to divert blame from the UK government's failures.

A World Health Organization report says China ``rolled out perhaps the most ambitious, agile \& aggressive disease containment effort in history''

We haven't even sorted out enough tests for NHS staff
\end{displayquote}

\subsection*{April 1st to 16th}

This section reviews April 1 up to and including April 16th 2020, before Trump issued his liberation tweets to Virginia, Michigan and Minnesota and a backlash against lockdown measures became apparent.

\begin{figure}
  \includegraphics[width=.85\textwidth]{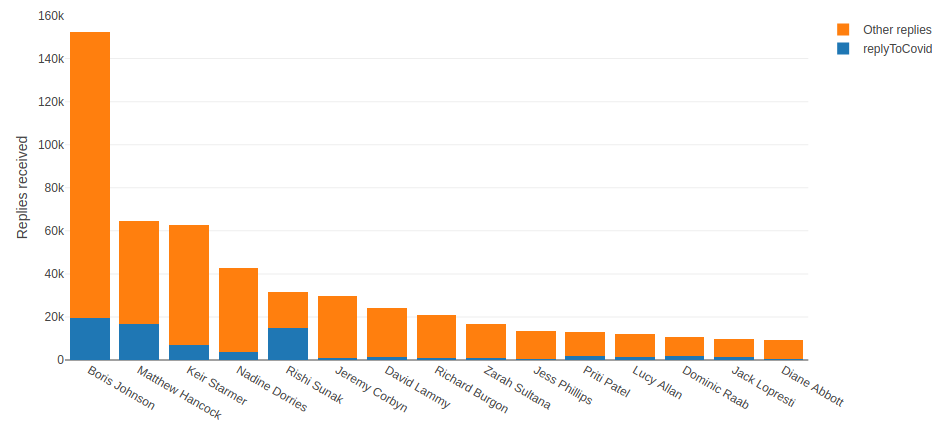}
  \caption{Number of replies with an explicit COVID-19 mention as a portion of all replies from April 1st - 16th 2020 inclusive.}
  \label{fig:apr-1-histo}
\end{figure}

As indicated by the table and the histogram depicted in Table~\ref{tab:apr-1} and Fig~\ref{fig:apr-1-histo} respectively, Boris Johnson's abuse level is extremely low as his illness takes a serious turn. Keir Starmer begins to attract attention in his new role as Labour leader, and the attention is much less abusive than Jeremy Corbyn tended to receive in the same role (10\% in the first half of 2019,\footnote{\small{\url{https://arxiv.org/pdf/1910.00920.pdf}}} but the tables shown here also show consistently high abuse levels for Mr Corbyn). Jeremy Corbyn begins to attract less attention by volume of replies on Twitter compared to others. Nadine Dorries attracts a higher abuse level than Matthew Hancock and Rishi Sunak for a tweet given below. 

The high abuse level toward Jack Lopresti during this period relates to his controversial opinion that churches should open for Easter. Example tweets are given below.

\begin{table}
\begin{tabular}{lrrrrrrr}
Name & Authored & replyTo & Covid & Abusive & \% Ab & \% Covid & Total Covid\\
\hline
\cellcolor{blue!15}Boris Johnson & 16 & 152,238 & 19,374 & 2,942 & 1.933 & 12.726 & 19.596\\
\cellcolor{blue!15}Matthew Hancock & 77 & 64,584 & 16,957 & 2,714 & 4.202 & 26.256 & 17.151\\
\cellcolor{red!15}Keir Starmer & 44 & 62,589 & 6,802 & 1,939 & 3.098 & 10.868 & 6.880\\
\cellcolor{blue!15}Nadine Dorries & 62 & 42,521 & 3,663 & 2,056 & 4.835 & 8.615 & 3.705\\
\cellcolor{blue!15}Rishi Sunak & 28 & 31,700 & 14,724 & 241 & 0.760 & 46.448 & 14.893\\
\cellcolor{red!15}Jeremy Corbyn & 29 & 29,548 & 1,085 & 1,445 & 4.890 & 3.672 & 1.097\\
\cellcolor{red!15}David Lammy & 73 & 24,254 & 1,203 & 713 & 2.940 & 4.960 & 1.217\\
\cellcolor{red!15}Richard Burgon & 73 & 21,084 & 776 & 1,291 & 6.123 & 3.681 & 0.785\\
\cellcolor{red!15}Zarah Sultana & 79 & 16,789 & 957 & 1,005 & 5.986 & 5.700 & 0.968\\
\cellcolor{red!15}Jess Phillips & 113 & 13,337 & 542 & 256 & 1.919& 4.064 & 0.548\\
\cellcolor{blue!15}Priti Patel & 22 & 13,063 & 1,785 & 252 & 1.929 & 13.665 & 1.805\\
\cellcolor{blue!15}Lucy Allan & 78 & 12,117 & 1,358 & 612 & 5.051 & 11.207 & 1.374\\
\cellcolor{blue!15}Dominic Raab & 41 & 10,701 & 1,823 & 203 & 1.897 & 17.036 & 1.844\\
\cellcolor{blue!15}Jack Lopresti & 14 & 9,736 & 1,385 & 1,655 & 16.999 & 14.226 & 1.401\\
\cellcolor{red!15}Diane Abbott & 66 & 9,486 & 513 & 221 & 2.330 & 5.408 & 0.519\\
\end{tabular}
\caption{MPs with greatest number of replies from from April 1 - 16 2020 inclusive. Cell colours indicate party membership; blue for Conservative, red for Labour.}
\label{tab:apr-1}
\end{table}

\subsubsection*{Hashtags}

\begin{figure}
  \includegraphics[width=.85\textwidth]{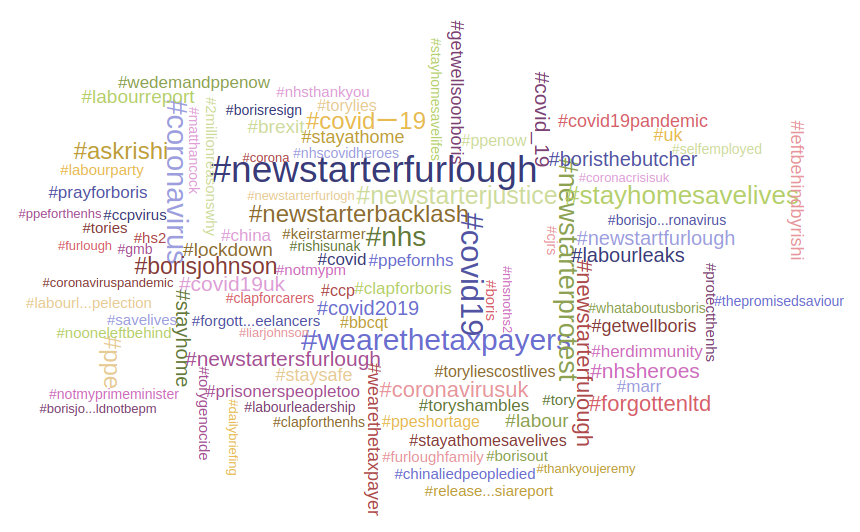}
  \caption{Top 100 hashtags in all replies sent to MPs - April 1st to April 16th 2020 inclusive.}
  \label{fig:apr-1-cloud}
\end{figure}

The hashtag cloud in Fig~\ref{fig:apr-1-cloud} shows that attention has begun to focus on the economic cost of the lockdown, as illustrated by the prominence of hashtags such as \#newstarterfurlough and \#wearethetaxpayers. Table~\ref{tab:apr-1-hash} gives counts for the top ten.

\begin{table}
\begin{tabular}{lr}
Hashtag & Count\\
\hline
\#newstarterfurlough & 30,511\\
\#covid19 & 8,923\\
\#wearethetaxpayers & 8,483\\
\#coronavirus & 6,849\\
\#nhs & 5,032\\
\#newstarterprotest & 4,282\\
\#stayhomesavelives & 3,619\\
\#newstarterjustice & 2,898\\
\#newstarterbacklash & 2,259\\
\#askrishi & 2,243\\
\end{tabular}
\caption{Hashtag counts in replies to MPs, April 1st - 16th.}
\label{tab:apr-1-hash}
\end{table}

\subsubsection*{Examples of tweets that attracted particularly abusive responses}

\url{https://twitter.com/JackLopresti/status/1247508135029411841} (18\% abuse, 6\% of all abuse sent to MPs between April 1st and 16th inclusive):

\begin{displayquote}
Open the churches for Easter – and give people hope \url{https://telegraph.co.uk/news/2020/04/06/open-churches-easter-give-people-hope/?WT.mc_id=tmg_share_tw…} via @telegraphnews
\end{displayquote}

\url{https://twitter.com/JackLopresti/status/1247894726486798342} (17\% abuse, 3\% of all abuse sent to MPs between April 1st and 16th inclusive):

\begin{displayquote}
Today I wrote to The Secretary of State @mhclg and also sent a copy of this letter to Secretary of State @DCMS to ask the Government to consider opening church doors on Easter Sunday for private prayer.
\end{displayquote}

\url{https://twitter.com/NadineDorries/status/1248329497897775105} (11\% abuse, 5\% of all abuse sent to MPs between April 1st and 16th inclusive - given that many people were dying, of a respiratory virus, it seemed tactless)

\begin{displayquote}
The boss is in a better place. Such a relief. The country can breathe again
\end{displayquote}

\url{https://twitter.com/RichardBurgon/status/1247248198932062208} (17\% abuse, 2\% of all abuse sent to MPs between April 1st and 16th inclusive - regarding his work as shadow justice secretary, regarded by some as mistimed considering the prime minister's health at the time).

\subsection*{April 17th to May 9th}

\begin{figure}
  \includegraphics[width=.85\textwidth]{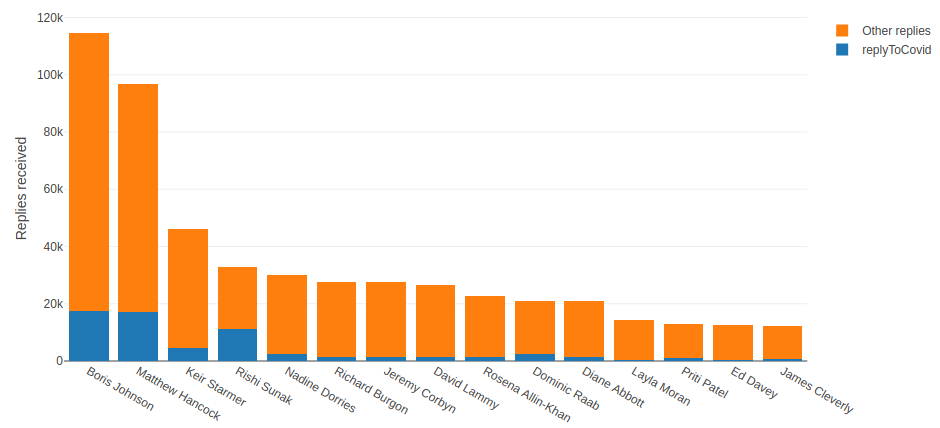}
  \caption{Number of replies with an explicit COVID-19 mention as a portion of all replies from April 17th - May 9th 2020 inclusive.}
  \label{fig:apr-17-histo}
\end{figure}

The histogram and table in Fig~\ref{fig:apr-17-histo} and Table~\ref{tab:apr-17} respectively reflect the unusual circumstances of this period. Boris Johnson did not return to work until April 26th, so the greater prominence of Matt Hancock on Twitter during this period, while the prime minister recuperated at Chequers, perhaps reflects this. Dominic Raab, in his role as acting prime minister, did not attract high attention levels on Twitter. Keir Starmer is now the most replied-to Labour politician on Twitter, but continues to attract low abuse levels.

\begin{table}
\begin{tabular}{lrrrrrrr}
Name & Authored & replyTo & Covid & Abusive & \% Ab & \% Covid & Total Covid\\
\hline
\cellcolor{blue!15}Boris Johnson & 40 & 114,506 & 17,419 & 3,744 & 3.270 & 15.212 & 17.646\\
\cellcolor{blue!15}Matthew Hancock & 104 & 96,712 & 17,017 & 4,175 & 4.317 & 17.596 & 17.238\\
\cellcolor{red!15}Keir Starmer & 55 & 46,202 & 4,668 & 1,286 & 2.783 & 10.103 & 4.729\\
\cellcolor{blue!15}Rishi Sunak & 50 & 32,702 & 11,042 & 421 & 1.287 & 33.766 & 11.186\\
\cellcolor{blue!15}Nadine Dorries & 79 & 30,093 & 2,499 & 1,006 & 3.343 & 8.304 & 2.532\\
\cellcolor{red!15}Richard Burgon & 105 & 27,760 & 1,445 & 2,536 & 9.135 & 5.205 & 1.464\\
\cellcolor{red!15}Jeremy Corbyn & 51 & 27,534 & 1,422 & 1,960 & 7.118 & 5.165 & 1.440\\
\cellcolor{red!15}David Lammy & 81 & 26,432 & 1,286 & 1,267 & 4.793 & 4.865 & 1.303\\
\cellcolor{red!15}Rosena Allin-Khan & 69 & 22,553 & 1,443 & 656 & 2.909 & 6.398 & 1.462\\
\cellcolor{blue!15}Dominic Raab & 63 & 20,988 & 2,344 & 584 & 2.783 & 11.168 & 2.374\\
\cellcolor{red!15}Diane Abbott & 136 & 20,966 & 1,307 & 574 & 2.738 & 6.234 & 1.324\\
\cellcolor{yellow!40}Layla Moran & 131 & 14,466 & 356 & 1,195 & 8.261 & 2.461 & 0.361\\
\cellcolor{blue!15}Priti Patel & 23 & 12,764 & 976 & 207 & 1.622 & 7.647 & 0.989\\
\cellcolor{yellow!40}Ed Davey & 84 & 12,425 & 472 & 1,618 & 13.022 & 3.799 & 0.478\\
\cellcolor{red!15}James Cleverly & 61 & 12,210 & 780 & 467 & 3.825 & 6.388 & 0.790\\
\end{tabular}
\caption{MPs with greatest number of replies from from April 17 - May 9 2020 inclusive. Cell colours indicate party membership; blue for Conservative, red for Labour, yellow for Liberal Democrat.}
\label{tab:apr-17}
\end{table}

\subsubsection*{Hashtags}

\begin{figure}
  \includegraphics[width=.85\textwidth]{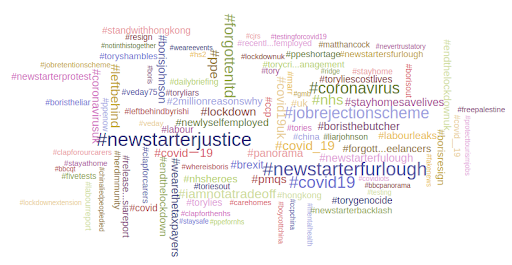}
  \caption{Top 100 hashtags in all replies sent to MPs - April 17th to May 4th 2020 inclusive.}
  \label{fig:apr-17-cloud}
\end{figure}

Later in the month, the economic focus continues, as shown by the hashtag cloud in Fig~\ref{fig:apr-17-cloud}. The majority of hashtags now appear to be critical, often economically focused but also including accusations of lying against China, Boris Johnson and Conservatives, and references to the shortage of personal protective equipment for medical workers. The distinct change in tone echoes events in the USA.\footnote{\small{E.g. \url{https://www.theguardian.com/global/video/2020/apr/16/armed-protesters-demand-an-end-to-michigans-coronavirus-lockdown-orders-video}}} Table~\ref{tab:apr-17-hash} gives counts for the top ten.

\begin{table}
\begin{tabular}{lr}
Hashtag & Count\\
\hline
\#newstarterjustice & 22,833\\
\#newstarterfurlough & 18,760\\
\#covid19 & 9,114\\
\#jobrejectionscheme & 6,351\\
\#coronavirus & 6,168\\
\#forgottenltd & 5,307\\
\#nhs & 3,535\\
\#iamnotatradeoff & 2,679\\
\#ppe & 2,553\\
\#leftbehind & 2,402\\
\end{tabular}
\caption{Hashtag counts in replies to MPs, April 17th - May 9th.}
\label{tab:apr-17-hash}
\end{table}

\subsubsection*{Examples of tweets that attracted particularly abusive responses}

The tweet receiving the most abusive response by volume also received a striking level of abuse by percentage; this one by Ed Davey.

\url{https://twitter.com/EdwardJDavey/status/1253882262715842560} (19\% abuse, 5\% of all abuse toward MPs for the period):

\begin{displayquote}
A pre-dawn meal today

Preparing for my first ever fast in the holy month of Ramadan

For Muslims doing Ramadan in isolation, you are not alone!

\#RamadanMubarak

\#LibDemIftar
\end{displayquote}

Two tweets by Richard Burgon also received a high level of abuse by volume and as a percentage:

\url{https://twitter.com/RichardBurgon/status/1252168331618123776} (16\% abuse, 4\% of all abuse for the period):

\begin{displayquote}
So many countries are doing much better than we are at tackling Coronavirus

Failure after failure is costing lives. We must speak out.

UK - 228 deaths per million people

Ireland - 116

Portugal - 67

Germany - 51

Canada - 39

South Korea - 5

China - 3

Australia - 3

New Zealand - 2
\end{displayquote}

\url{https://twitter.com/RichardBurgon/status/1256184006741241857} (13\% abuse, 2\% of all abuse for the period):

\begin{displayquote}
The UK has had more Coronavirus deaths than the following countries COMBINED  

Belgium

Germany

Netherlands

Switzerland

Ireland

Portugal

Romania

Poland

Austria

Denmark

Ukraine

Finland

Norway

Greece

The UK has 67m people. The others around 270m people.
\end{displayquote}

The following tweets also attracted high levels of abuse by volume:

\url{https://twitter.com/jeremycorbyn/status/1253341601599852544} (11\% abusive replies, 2\% of abuse for the period - this was also St George's Day, so perceived as evidence of anti-English sentiment, as in the following paraphrased replies for example: ``@jeremycorbyn So nothing about St George's day then? Ah, that's because we are English, the country you wanted to run but hate with a vengeance. And you wonder why you suffered such a huge defeat at the election'' and ``@jeremycorbyn So no mention of St. George's day then? You utter cretin.''):

\begin{displayquote}
Ramadan Mubarak to all Muslims in Islington North, all across the UK and all over the world.
\end{displayquote}

\url{https://twitter.com/Simon4NDorset/status/1255043717045596160} (7\% abusive replies, 2\% of abuse for the period):

\begin{displayquote}
I’m afraid @piersmorgan is not acting as a journalist. As a barrack room lawyer? Yes. As a saloon bar bore? Yes. As a bully? Yes. As a show off? Undoubtedly.  He is not a seeker after truth: he’s a male chicken
\end{displayquote}

\subsubsection*{Non-MP accounts}

\begin{figure}
  \includegraphics[width=.85\textwidth]{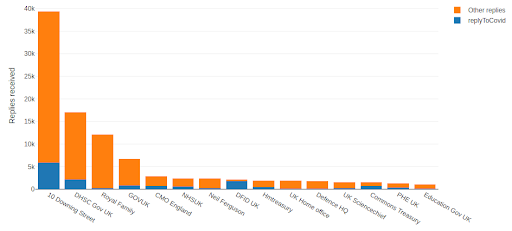}
  \caption{Covid-related replies to other accounts, Apr 17th - May 9th.}
  \label{fig:apr-17-histo-other}
\end{figure}

From this time period, we began also to collect data for a set of government accounts and other accounts relevant to the epidemic. In this period, as shown in Table~\ref{tab:apr-17-other} and Fig~\ref{fig:apr-17-histo-other}, Neil Ferguson, a medical expert formerly included in the Scientific Advisory Group for Emergencies (SAGE), who advise the UK government, received more attention than he goes on to receive in the following period, and a high level of abuse. Mr Ferguson resigned from SAGE during this period, following publicity surrounding his lockdown violation. Note that Neil Ferguson was also targeted by conspiracy theorists.\footnote{\small{\url{https://thefederalist.com/2020/03/26/the-scientist-whose-doomsday-pandemic-} \url{model-predicted-armageddon-just-walked-back-the-apocalyptic-predictions/}}} Chief medical officer of England Chris Whitty (``CMO\_England'') and chief scientific advisor Patrick Vallance (``uksciencechief'') both appear in the table, but receive only a fraction of the total and Covid-related attention that 10 Downing Street receives.

\begin{table}
\begin{tabular}{lrrrrrrr}
Name & Authored & replyTo & Covid & Abusive & \% Ab & \% Covid & Total Covid\\
\hline
10 Downing Street & 208 & 39,348 & 5,852 & 933 & 2.371 & 14.872 & 42.526\\
DHSC Gov UK & 288 & 17,071 & 2,188 & 216 & 1.265 & 12.817 & 15.900\\
Royal Family & 66 & 12,076 & 261 & 67 & 0.555 & 2.161 & 1.897\\
GOV UK & 66 & 6,703 & 879 & 176 & 2.626 & 13.114 & 6.388\\
CMO Eng. (Whitty) & 6 & 2,943 & 734 & 33 & 1.121 & 24.941 & 5.334\\
NHS UK & 97 & 2,458 & 588 & 21 & 0.854 & 23.922 & 4.273\\
Neil Ferguson & 1 & 2,394 & 266 & 266 & 11.111 & 11.111 & 1.933\\
HM Treasury & 48 & 1,981 & 503 & 13 & 0.656 & 25.391 & 3.655\\
UK Home Office & 55 & 1,894 & 89 & 38 & 2.006 & 4.699 & 0.647\\
Defence HQ & 146 & 1,862 & 88 & 8 & 0.430 & 4.726 & 0.639\\
UK Sci C. (Vallance) & 8 & 1,587 & 294 & 13 & 0.819 & 18.526 & 2.136\\
Commons Treasury & 34 & 1,511 & 680 & 8 & 0.529 & 45.003 & 4.942\\
PHE UK & 58 & 1,373 & 349 & 7 & 0.510 & 25.419 & 2.536\\
Foreign Office & 72 & 963 & 41 & 14 & 1.454 & 4.258 & 0.298\\
UK Parliament & 25 & 743 & 137 & 8 & 1.077 & 18.439 & 0.996\\
\end{tabular}
\caption{Statistics for other accounts from from April 17th - May 9th 2020 inclusive.}
\label{tab:apr-17-other}
\end{table}

\subsection*{May 10th to 25th}

\begin{figure}
  \includegraphics[width=.85\textwidth]{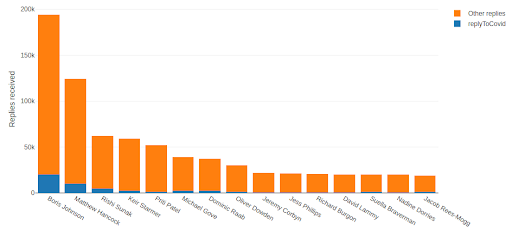}
  \caption{Number of replies with an explicit COVID-19 mention as a portion of all replies from May 10th - 25th 2020 inclusive.}
  \label{fig:may-10-histo}
\end{figure}

In Table~\ref{tab:may-10} and Fig~\ref{fig:may-10-histo} we see a return to a high level of focus on Boris Johnson, with other senior ministers also prominent. Of Labour politicians, only Keir Starmer attracts notable attention, with others much further down the list. Higher levels of abuse are received by ministers who defended Dominic Cummings on Twitter; Matthew Hancock, Oliver Dowden and Michael Gove. Boris Johnson also receives more abuse than he did in the previous period. Example tweets are given below of ministers defending Mr Cummings.

\begin{table}
\begin{tabular}{lrrrrrrr}
Name & Authored & replyTo & Covid & Abusive & \% Ab & \% Covid & Total Covid\\
\hline
\cellcolor{blue!15}Boris Johnson & 42 & 194,083 & 20,109 & 9,543 & 4.917 & 10.361 & 27.014\\
\cellcolor{blue!15}Matthew Hancock & 59 & 124,070 & 10,031 & 7,905 & 6.371 & 8.085 & 13.475\\
\cellcolor{blue!15}Rishi Sunak & 21 & 62,304 & 5,001 & 2,292 & 3.679 & 8.027 & 6.718\\
\cellcolor{red!15}Keir Starmer & 47 & 59,239 & 2,501 & 2,063 & 3.483 & 4.222 & 3.360\\
\cellcolor{blue!15}Priti Patel & 23 & 51,989 & 1,364 & 2,184 & 4.201 & 2.624 & 1.832\\
\cellcolor{blue!15}Michael Gove & 9 & 39,171 & 2,660 & 2,791 & 7.125 & 6.791 & 3.573\\
\cellcolor{blue!15}Dominic Raab & 32 & 37,433 & 2,318 & 2,135 & 5.704 & 6.192 & 3.114\\
\cellcolor{blue!15}Oliver Dowden & 21 & 29,929 & 1,015 & 2,446 & 8.173 & 3.391 & 1.364\\
\cellcolor{red!15}Jeremy Corbyn & 29 & 22,167 & 731 & 1,513 & 6.825 & 3.298 & 0.982\\
\cellcolor{red!15}Jess Phillips & 102 & 21,449 & 545 & 541 & 2.522 & 2.541 & 0.732\\
\cellcolor{red!15}Richard Burgon & 74 & 20,483 & 588 & 1,466 & 7.157 & 2.871 & 0.790\\
\cellcolor{red!15}David Lammy & 59 & 20,363 & 599 & 837 & 4.110 & 2.942 & 0.805\\
\cellcolor{blue!15}Suella Braverman & 5 & 20,199 & 1,019 & 716 & 3.545 & 5.045 & 1.369\\
\cellcolor{blue!15}Nadine Dorries & 29 & 20,014 & 511 & 740 & 3.697 & 2.553 & 0.686\\
\cellcolor{blue!15}Jacob Rees-Mogg & 7 & 19,151 & 1,011 & 1,439 & 7.514 & 5.279 & 1.358\\
\end{tabular}
\caption{MPs with greatest number of replies from from May 10 - May 25 2020 inclusive. Cell colours indicate party membership; blue for Conservative, red for Labour.}
\label{tab:may-10}
\end{table}

\subsubsection*{Hashtags}

\begin{figure}
  \includegraphics[width=.85\textwidth]{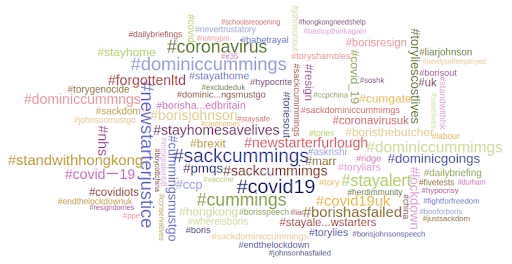}
  \caption{Top 100 hashtags for all replies sent to MPs - May 10th - 25th 2020 inclusive.}
  \label{fig:may-10-cloud}
\end{figure}

Hashtags in Fig~\ref{fig:may-10-cloud} show a high degree of negative attention focused on Dominic Cummings, whilst continued attention on the economic plight of new starters is also in evidence. Table~\ref{tab:may-10-hash} gives counts for the top ten.

\begin{table}
\begin{tabular}{lr}
Hashtag & Count\\
\hline
\#covid19 & 7,352\\
\#sackcummings & 6,054\\
\#newstarterjustice & 5,446\\
\#dominiccummings & 5,004\\
\#coronavirus & 4,412\\
\#cummings & 4,093\\
\#stayalert & 3,327\\
\#dominiccummimgs & 2,727\\
\#borishasfailed & 2,265\\
\#standwithhongkong & 2,229\\
\end{tabular}
\caption{Hashtag counts in replies to MPs, May 10th - 25th inclusive.}
\label{tab:may-10-hash}
\end{table}

\subsubsection*{Examples of tweets that attracted particularly abusive responses}

The four tweets to attract the highest volumes of abuse were all by cabinet members defending Dominic Cummings. Abuse levels are high but not the highest we have seen. There was clearly a high level of attention on the issue however.

\url{https://twitter.com/MattHancock/status/1264162359733555202} (7\% abuse, 7\% of abuse for the period):

\begin{displayquote}
I know how ill coronavirus makes you. It was entirely right for Dom Cummings to find childcare for his toddler, when both he and his wife were getting ill.
\end{displayquote}

\url{https://twitter.com/OliverDowden/status/1264221876374646786} (8\% abuse, 5\% of abuse for the period):

\begin{displayquote}
Dom Cummings followed the guidelines and looked after his family. End of story.
\end{displayquote}

\url{https://twitter.com/michaelgove/status/1264126108733186050} (8\% abuse, 5\% of abuse for the period):

\begin{displayquote}
Caring for your wife and child is not a crime
\end{displayquote}

\url{https://twitter.com/MattHancock/status/1264975804208947208} (9\% abuse, 4\% of abuse for the period):

\begin{displayquote}
Dom Cummings was right today to set out in full detail how he made his decisions in very difficult circumstances. Now we must move on, fight this dreadful disease and get our country back on her feet
\end{displayquote}

As a percentage of replies, the most notable tweet was the following:

\url{https://twitter.com/HenrySmithUK/status/1263394101002674176} (13\% abuse, 2\% of abuse for the period):

\begin{displayquote}
Not that I should be surprised by the lazy left but interesting how work-shy socialist and nationalist MPs tried to keep the remote Parliament going beyond 2 June.
\end{displayquote}

\subsubsection*{Non-MP accounts}

\begin{figure}
  \includegraphics[width=.85\textwidth]{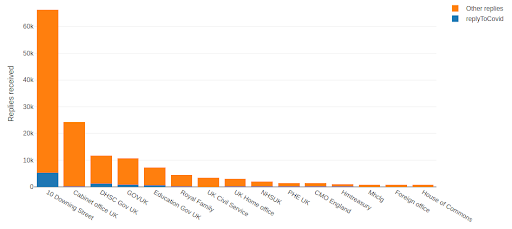}
  \caption{Covid-related replies to other accounts as a portion of all replies received, May 10th - 25th inclusive.}
  \label{fig:may-10-histo-other}
\end{figure}

Dominic Cummings does not have a clearly labelled and verified Twitter account, though the account ``OdysseanProject'', rumoured to be his, does show elevated attention in this period, and some abuse, though not sufficient to appear in the top 15 non-MP accounts shown in Fig~\ref{fig:may-10-histo-other} and Table~\ref{tab:may-10-other}. It is unlikely that many Twitter users are aware of this anonymous account (and indeed, our information may be incorrect!) However the extent of the controversy around Mr Cummings' lockdown violation shows itself better in responses to MPs defending his actions, and in the use of hashtags, as shown above.

\begin{table}
\begin{tabular}{lrrrrrrr}
Name & Authored & replyTo & Covid & Abusive & \% Ab & \% Covid & Total Covid\\
\hline
10 Downing Street & 146 & 66,266 & 5,190 & 2,486 & 3.752 & 7.832 & 50.536\\
Cabinet Office UK & 36 & 24,239 & 297 & 627 & 2.587 & 1.225 & 2.892\\
DHSC Gov UK & 182 & 11,714 & 1,208 & 149 & 1.272 & 10.312 & 11.762\\
GOVUK & 35 & 10,767 & 777 & 432 & 4.012 & 7.216 & 7.566\\
Education Gov UK & 102 & 7,232 & 561 & 71 & 0.982 & 7.757 & 5.463\\
Royal Family & 32 & 4,515 & 192 & 25 & 0.554 & 4.252 & 1.870\\
UK Civil Service & 91 & 3,535 & 49 & 39 & 1.103 & 1.386 & 0.477\\
UK Home Office & 50 & 2,947 & 133 & 74 & 2.511 & 4.513 & 1.295\\
NHS UK & 94 & 1,985 & 286 & 26 & 1.310 & 14.408 & 2.785\\
PHE UK & 53 & 1,460 & 222 & 10 & 0.685 & 15.205 & 2.162\\
CMO Eng. (Whitty) & 2 & 1,450 & 277 & 18 & 1.241 & 19.103 & 2.697\\
HM Treasury & 49 & 1,069 & 243 & 6 & 0.561 & 22.732 & 2.366\\
MHCLG & 45 & 895 & 49 & 15 & 1.676 & 5.475 & 0.477\\
Foreign Office & 50 & 838 & 46 & 4 & 0.477 & 5.489 & 0.448\\
House of Commons & 71 & 818 & 101 & 4 & 0.489 & 12.347 & 0.983\\
\end{tabular}
\caption{Statistics for other accounts from from May 10th - 25th 2020 inclusive.}
\label{tab:may-10-other}
\end{table}

\subsection*{Conspiracy theories}

This report\footnote{\small{\url{http://moonshotcve.com/covid-19-conspiracy-theories-hate-speech-twitter/}}} from Moonshot CVE was used as a guide to the overall conspiracy landscape within COVID-19. The areas they highlight are anti-Chinese feeling/conspiracy theory, theories that link the virus to a Jewish plot, theories that link the virus to an American plot, generic ``deep state'' and 5g-based theories and general theories that the virus is a plot or hoax. Further search areas were then added as controls. The three controls appear at the bottom of Table~\ref{tab:conspiracies}; ``\#endthelockdown'' and close variants, ``\#newstarterfurlough'' and variants, and ``\#stayhomesavelives'', in order to contrast volumes with anti-lockdown feeling, the leading economy-related campaign and pro-lockdown feeling respectively.

The table shows substantial evidence of ill-feeling toward China. Classic conspiracy theories are in evidence but numbers of mentions are low (though note that most of the 183 mentions of ``NWO'' (``new world order'') are now COVID-19-related, suggesting opportunistic incorporation of COVID-19 into existing mythologies). There is considerable evidence of some Twitter users not believing in the virus, and that numbers of mentions to this effect are within one order of magnitude of the popular ``stay home save lives''. Yet all are surpassed by the theme of economic support for those not in established employment.

\begin{table}
\resizebox{\textwidth}{!}{%
\begin{tabular}{lrrr}
\textbf{Search terms (in all replies to MPs, not case-sensitive)} & \textbf{\# tweets} & \textbf{\# abusive} & \textbf{\% abusive}\\
\hline
``NukeChina'' OR ``BombChina'' OR ``DeathtoChina'' OR ``DestroyChina'' & 31 & 2 & 6.452\\
OR ``Nuke China'' OR ``Bomb China'' OR ``Death to China'' OR & \\
``Destroy China'' OR ``\#nukechina'' OR ``\#bombchina'' OR \\ ``\#deathtochina'' OR ``\#destroychina'' & \\
\hline
``ccpvirus'' OR ``chinaliedpeopledied'' OR ``ccp virus'' OR & 2,463 & 38 & 1.543\\
``china lied people died'' OR ``\#ccpvirus'' OR ``\#chinaliedpeopledied'' &\\
\hline
``\#chinesevirus'' OR ``chinesevirus'' OR ``chinese virus'' & 2,208 & 119 & 5.389\\
\hline
``Soros Virus'' OR ``Israel Virus'' OR `` NWO Virus'' OR ``SorosVirus'' OR  & 3 (all & 0 & 0\\
``IsraelVirus'' OR `` NWOVirus'' OR ``\#sorosvirus'' OR ``\#israelvirus'' OR & ``NWO\\
``\#nwovirus'' & virus'')\\
\hline
``\#nwo'' & 183 & 8 & 4.372\\
\hline
``Gates Virus'' OR ``CIA Virus'' OR ``America Virus'' OR ``GatesVirus'' OR  & 69 & 2 & 2.899\\
``CIAVirus'' OR ``AmericaVirus'' OR ``\#gatesvirus'' OR ``\#ciavirus'' OR  & \\
``\#americavirus'' & \\
\hline
``deepstatevirus'' OR ``deep state virus'' OR ``\#deepstatevirus'' & 2 & 0 & 0\\
\hline
``\#5gcoronavirus'' & 53 & 1 & 1.877\\
\hline
``\#5gkills'' & 70 & 0 & 0\\
\hline
``coronahoax'' OR ``corona hoax'' OR ``hoax virus'' OR ``fake virus'' OR  & 415 & 61 & 14.699\\
``hoaxvirus'' OR ``fakevirus'' OR ``\#coronahoax'' OR ``\#hoaxvirus'' OR & \\
``\#virushoax'' OR ``\#fakevirus'' & \\
\hline
``corona bollocks'' OR ``coronabollocks'' OR ``corona bollox'' OR ``coronabollox''  & 250 & 19 & 7.600\\
OR ``\#coronabollocks'' OR ``\#coronabollox'' & \\
\hline
``plandemic'' OR ``scamdemic'' OR ``\#plandemic'' OR ``\#scamdemic'' OR  & 1,178 & 64 & 5.433\\
``\#fakepandemic'' OR ``\#whereisthepandemic'' OR ``\#plandemic2020'' & \\
\hline
``\#filmyourhospital'' OR ``\#emptyhospitals'' & 59 & 5 & 8.475\\
\hline
``end the lockdown'' OR ``endthelockdown'' OR ``end lockdown'' OR  & 5,506 & 272 & 4.940\\
``endlockdown'' OR ``end lock down'' OR ``\#endlockdown'' OR & \\
``\#endthelockdown'' OR ``\#lockdownend'' OR ``\#endthislockdown'' OR & \\
``\#endthelockdownuk'' OR ``\#endthelockdownnow'' & & \\
\hline
``newstarterfurlough'' OR ``new starter furlough'' OR ``new starter justice'' OR & 55,593 & 243 & 0.437\\
OR ``newstarterjustice'' OR hashtag\_string:``\#newstarterfurlough'' OR & \\
``newstarterjustice'' OR ``\#newstarterfurlough'' OR ``\#newstarterjustice'' & \\
\hline
``stayhomesavelives'' OR ``stay home save lives'' OR ``\#stayhomesavelives'' OR & 20,538 & 1222 & 5.950\\
``\#stayathomesavelives'' OR ``\#stayathomeandsavelives'' OR & \\
``\#stayhomestaysafe'' OR ``\#stayhomeandstaysafe'' OR & \\
``\#stayathomestaysafe'' OR ``\#stayathomeandstaysafe'' OR ``\#stayathome'' & & &\\
OR ``\#stayhome'' & & &\\
\hline
\end{tabular}%
}
\caption{Mention count of conspiracy-related strings, alongside controls (last three rows), in all replies to MPs, Feb 7th to May 25th inclusive}
\label{tab:conspiracies}
\end{table}

\section*{Conclusion}

The crisis has led to elevated engagement with UK politicians by the public, and we have seen that this may be more positive and less abusive than the dialogue at other times. The leading hashtag campaign of the period, ``\#newstarterfurlough'', is associated with a remarkably low level of abuse (\textless0.5\% of replies) despite being a complaint hashtag. The surge of attention on Boris Johnson during his illness was substantially lower in abuse than his previous levels. Receiving more tweets mentioning the virus is associated with receiving lower levels of abuse. It may be that the crisis is leading to different people engaging with politicians than usually do, who are less inclined to verbally abuse them than those that usually occupy the space.

Yet the usual, more uncivil contingent remains active on Twitter, with politicians receiving abuse for particular topics, that may or may not be COVID-19-related, in much the same manner as they did before. Tweets from MPs expressing positive engagement with the Muslim community have been met with hostile and abusive responses, and hashtags associating the virus with China have an elevated likelihood of abuse, continuing an already noted pattern~\cite{gorrell2018twits, gorrell2019race} that racism and xenophobia are associated with particularly abusive tweets. Previous work has also described a substantial presence of overt Islamophobia in dialogue with MPs~\cite{gorrell2019race}. Xenophobia has not gone away, and indeed has found new fuel in the crisis.

In terms of responses to the handling of the crisis, feelings run high on both sides. Elevated levels of abuse are associated with hashtags supporting lockdown as well as those opposing it. Labour politicians in favour of stricter measures have received abusive responses, as have Conservative politicians defending Dominic Cummings' lockdown violation. New Labour leader Keir Starmer receives less abuse than his predecessor Jeremy Corbyn.

1,902 replies to MPs were found containing hashtags or terms that refute the existence of the virus (e.g. \#coronahoax, \#coronabollocks, 0.04\% of a total 4.7 million replies, or 9\% of the number of mentions of "stay home save lives" and variants). Evidence of some members of the public believing in COVID-19 conspiracy theories was also found. The high prevalence of disbelieving the existence of the virus is a cause for concern.


\begin{backmatter}




\bibliographystyle{bmc-mathphys} 
\bibliography{covid-twitter-mp-abuse-white-paper}      

\end{backmatter}
\end{document}